\documentclass[10pt,conference]{IEEEtran}
\IEEEoverridecommandlockouts

\usepackage{cite}
\usepackage{amsmath,amssymb,amsfonts}
\usepackage{algorithmic}
\usepackage{graphicx}
\usepackage{textcomp}
\usepackage{xcolor}
\usepackage{multirow} % For multi-row cells in the header
\usepackage{booktabs} % For professional-looking tables (\toprule, \midrule, \bottomrule)
\newcommand{\increase}[2]{\textbf{#1}$\uparrow$#2\%}
\usepackage{amsmath}
\def\BibTeX{{\rm B\kern-.05em{\sc i\kern-.025em b}\kern-.08em
    T\kern-.1667em\lower.7ex\hbox{E}\kern-.125emX}}
\usepackage{booktabs} % For professional-looking tables (\toprule, \midrule, \bottomrule)
\usepackage{amsmath}  % For math symbols like \uparrow and \downarrow
\newcommand{\nochange}[1]{#1}

\newcommand{\decrease}[2]{\textbf{#1}$\downarrow$#2\%}

\newcommand{\newitem}[1]{\textbf{#1}}
\begin{document}
\author{
    \IEEEauthorblockN{Yueke Zhang}
    \IEEEauthorblockA{
    \textit{Vanderbilt University}\\
    yueke.zhang@vanderbilt.edu}
\and
    \IEEEauthorblockN{Yifan Zhang}
    \IEEEauthorblockA{
    \textit{Vanderbilt University}\\
    yifan.zhang@vanderbilt.edu}
\and
    \IEEEauthorblockN{Kevin Leach}
    \IEEEauthorblockA{
    \textit{Vanderbilt University}\\
    kevin.leach@vanderbilt.edu}
\and
    \IEEEauthorblockN{Yu Huang}
    \IEEEauthorblockA{
    \textit{Vanderbilt University}\\
    yu.huang@vanderbilt.edu}
}
% --- Custom commands for formatting the performance changes ---

\title{CodeGrad: Integrating Multi-Step Verification with Gradient-Based LLM Refinement
}

\maketitle

\begin{abstract}
While Large Language Models (LLMs) have demonstrated remarkable capabilities in code generation, they often produce solutions that lack guarantees of correctness, robustness, and efficiency. This limitation is particularly acute in domains requiring strict constraints. CodeGrad introduces a principled framework that integrates rigorous verification techniques directly into an iterative LLM-based generation loop. It uniquely treats code as a differentiable variable, converting structured feedback and mathematical constraints into a textual pseudo-gradient. This gradient guides the model to iteratively refine solutions, ensuring they are not only functional but also robust and mathematically justified.

We evaluate CodeGrad on the HumanEval, HumanEval+, and LiveCodeBench benchmarks. Our implementation outperforms strong baselines, achieving an absolute improvement of up to 27\% on HumanEval and a 41\% relative improvement on the challenging LiveCodeBench V6. StructuredGrad generates mathematically justified code that is robust and efficient, paving the way for reliable AI-assisted software development in high-stakes applications.
\end{abstract}


\section{Introduction}
Automated code generation is a transformative technology in software engineering and AI, accelerating development, reducing human error, and making programming more accessible~\cite{svyatkovskiy2020intellicode,becker2023programming,li2022competition}. Large Language Models (LLMs) have demonstrated remarkable capabilities in generating code from natural language descriptions, solving competitive programming problems, and even synthesizing complex algorithms~\cite{gu2023llm,wang2023review,liu2023your}. 

%While early explorations into LLM-based code synthesis and competitive programming centered on one-shot or few-shot generation~\cite{xiong2024hlspilot, ye2025uncovering, sagodi2024methodology}, the research frontier has shifted toward augmentation with feedback mechanisms. These include test-driven development loops and reinforcement learning from human feedback (RLHF)~\cite{shen2024policy, gehring2024rlef}. While these methods can produce syntactically correct and functional code, they often struggle with edge cases, fail to generalize beyond seen examples, and lack mechanisms for systematic improvement~\cite{wang2023review,liu2023your,li2022competition}.  While these methods can produce syntactically correct and functional code, they exhibit limitations in handling edge cases, fail to generalize robustly, and lack mechanisms for systematic improvement~\cite{liu2024exploring,zhang2025llm,ouyang2023llm}.

Early explorations in LLM-based code synthesis centered on one-shot or few-shot generation~\cite{xiong2024hlspilot, ye2025uncovering, sagodi2024methodology}, the recent research has shifted to augmentation with feedback mechanisms like test-driven development and RLHF~\cite{shen2024policy, gehring2024rlef}. Despite these methods can produce syntactically correct and functional code, the resulting code, while often functional, frequently struggles with edge cases, fails to generalize beyond seen examples, and lacks mechanisms for systematic improvement~\cite{wang2023review, liu2023your, li2022competition, liu2024exploring, zhang2025llm, ouyang2023llm}.

%On the other hand, traditional formal methods in software engineering offer mathematical rigor and correctness guarantees but are often too rigid, require specialized expertise, and do not scale well to the open-ended, natural language-driven tasks tackled by LLMs~\cite{liu1995limitations,woodcock2009formal}. This presents a clear research gap: LLMs offer flexibility but lack formal rigor, while formal methods provide rigor but lack flexibility~\cite{agarwal2024copilot,fakhoury2024llm}. The challenge, therefore, is to develop a principled, iterative framework that synergistically combines the generative power of LLMs with the verifiable guarantees of formal methods. However, previous works has prove the successfully use a natural language based formal approach to contain the behaviour for LLM model~\cite{cao2025informal,ospanov2025apollo,hassan2024llm,jha2023dehallucinating}. 

Conversely, traditional formal methods in software engineering provide mathematical rigor and verifiable guarantees of correctness. However, their application is often hindered by their rigidity, the need for specialized expertise, and their poor scalability, particularly for the open-ended, natural-language tasks where LLMs excel~\cite{liu1995limitations,woodcock2009formal}. This fundamental tension between the flexibility of LLMs and the rigor of formal methods creates a distinct research gap~\cite{agarwal2024copilot,fakhoury2024llm}. The central challenge, therefore, is to develop a principled framework that synergistically combines the generative power of LLMs with the verifiable guarantees of formal methods~\cite{zhang2024fusion}. Encouragingly, recent studies have demonstrated the potential of using natural language as a proxy for formal reasoning to guide and constrain LLM behavior, suggesting a viable path toward this integration~\cite{cao2025informal,ospanov2025apollo,hassan2024llm,jha2023dehallucinating,morishita2024enhancing}.

\begin{figure*}
\centering
\includegraphics[width=0.9\textwidth]{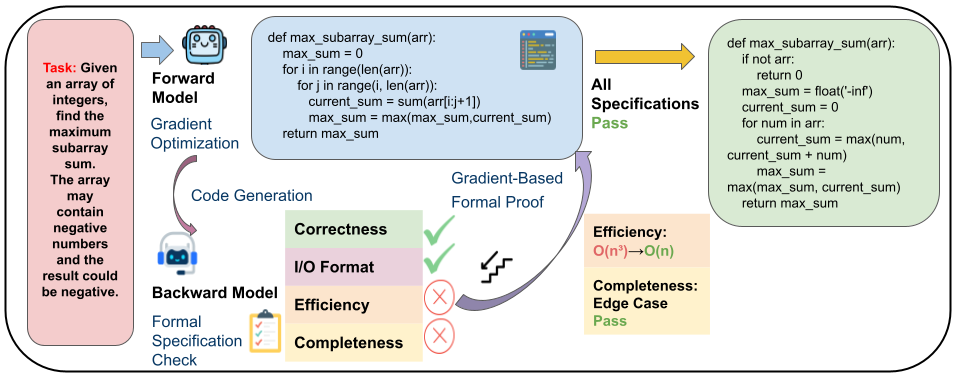}
\caption{Overview of the Formal-Gradient Code Generation loop on the “maximum-subarray” task. First, the forward model proposes an $O(n^3)$ triple-loop solution. The backward model then acts as a formal critic, analyzing the code to identify flaws—in this case, computational inefficiency and a failure to handle the all-negative numbers edge case. This structured critique is converted into a textual pseudo-gradient that directs refinement. Guided by this gradient and a formal efficiency invariant, the optimizer produces a new, correct, and highly efficient $O(n)$  solution. The loop terminates under one of two conditions: either the generated code is accepted upon successful formal verification of its accompanying proof, or the maximum number of iterations is reached.}
\label{fig:hook}
\vspace{-0.4cm}
\end{figure*}

In this work, we introduce \textbf{CodeGrad}, a novel framework integrating formal methods into an iterative, LLM-driven generation pipeline. As illustrated in Figure~\ref{fig:hook}, our approach treats code as a differentiable variable. This allows us to iteratively refine code by generating a textual pseudo-gradient—a structured set of corrections—derived from rigorous formal analysis. Rather than a numeric vector, this gradient consists of specific, textual edits that steer the code toward an optimal state. At each iteration, a backward engine, acting as a formal critic, evaluates the code against specifications for correctness, efficiency, and constraint compliance. The resulting critique is converted into the textual pseudo-gradient, which guides a forward engine to produce an improved candidate. To ensure rigor, a formal methods optimizer augments the update, enforcing explicit invariants. Crucially, it requires each revision to have a natural language “proof” of its correctness. This verification ensures solutions are not merely functional but robust and formally justified, guiding the process toward verifiably correct code. The contributions of this paper are:
%At each iteration, a backward engine evaluates the generated code, providing structured feedback on correctness, efficiency, and constraint compliance. This feedback is then propagated as a gradient to a forward engine, which produces an improved code candidate. We augment this process with a formal methods optimizer that enforces explicit invariants and requires each code update to be accompanied by a natural language ``proof'' of its correctness, ensuring that solutions are not only functional but also robust and formally justified.

%In this work, we introduce \textbf{CodeGrad}, a novel framework that integrates formal methods into the LLM-driven code generation pipeline to guide iterative refinement. As illustrated in Figure 1, our approach operationalizes this by treating code as a differentiable variable, enabling the use of textual pseudo-gradients to systematically improve solutions based on feedback from a formal critic. At each iteration, a backward engine evaluates the generated code, providing structured feedback on correctness, efficiency, and compliance with problem constraints. This feedback is then propagated as gradients, guiding the forward engine to produce improved code in subsequent iterations. We augment this process with a formal methods optimizer that enforces explicit invariants and requires each code update to be accompanied by a natural language “proof” of correctness.

%Inspired by these complementary strengths, we present CodeGrad, a principled, iterative refinement framework that bridges this gap.  our approach treats code as a differentiable variable within a feedback loop. A forward model generates a solution, which a backward critic then evaluates to produce structured feedback. This feedback is converted into a textual pseudo-gradient, guiding the next generation cycle toward a more optimal and formally justified solution.

\begin{enumerate}
    \item We present CodeGrad, the first framework to integrate formal methods to guide an LLM-based code generation loop using textual pseudo-gradients for iterative refinement.
    \item We conduct extensive evaluations on the HumanEval, HumanEval+, and LiveCodeBench benchmarks. Our results demonstrate the framework's effectiveness, achieving performance increase of up to 27.2\% on HumanEval and a improvement of 40.6\% on the lastest LiveCodeBench V6 benchmark over baselines.
    \item We analyze the framework's performance across various problem categories and difficulty levels, revealing its particular efficacy in complex algorithmic domains where baseline models completely fail. 
\end{enumerate}

%By establishing a method for generating code that is not only functional but also robust and formally justified, CodeGrad paves the way for more reliable AI-assisted software development. T

\section{Related Work}
Research into improving the reliability of LLM-generated code has evolved from enhancing intrinsic reasoning to employing sophisticated, external feedback loops~\cite{hao2024llm,zhang2024llm}. Early efforts focused on improving the step-by-step reasoning of models, but recent progress has been made with iterative refinement~\cite{zhou2024isr,abbasi2024believe}. Reflexion established a paradigm where an agent critiques its own output to guide subsequent improvements~\cite{shinn2023reflexion}. In the code generation domain, this principle was adapted into test-driven methods where feedback was derived from unit test outcomes, as seen in CODET and CODERL~\cite{le2022coderl,chen2022codet}. More recently, this feedback mechanism has been formalized through novel approaches like TextGrad, which conceptualizes natural language critiques as pseudo-gradients~\cite{yuksekgonul2024textgrad}. Textgrad use the model toward a correct solution by treating textual corrections as differentiable signals within the discrete space of code. Our work builds upon this trajectory, integrating gradient-style feedback with formal methods.

While TextGrad formalizes the general feedback mechanism, a parallel line of research has explored integrating principles from formal methods to better constrain LLM behavior. A key challenge is the difficulty and low accuracy of having LLMs generate code in a strict formal language~\cite{cao2025informal}. Consequently, recent work has focused on using natural language as a medium for formal reasoning. For instance, Formal-LLM has been shown to successfully guide the planning process for agents by preventing the generation of invalid or unsuccessful plans~\cite{li2024formal}. Other approaches have used formal methods to facilitate the self-monitoring of LLM responses or leveraged LLMs to formulate planning tasks as solvable optimization problems~\cite{jha2023dehallucinating,hao2024llm}.
%Also, existing approach has used formal methods to detect autonomous systems and sketch a novel self-monitoring errors in the LLM response automatically~\cite{jha2023dehallucinating}.  LLMFP, a general-purpose framework that leverages LLMs to capture key information from planning problems and formally formulate and solve them as optimization problems from scratch, with no task-specific examples needed~\cite{hao2024llm}.

\section{Methodology}
\label{sec:method}

\subsection{Problem Definition}
We define our problem as automatically generate a computer program that not only fulfills a natural language specification but also adheres to a set of predefined formal constraints. Formally, given a natural language task description $T$, our objective is to discover a program $P$ from the space of all possible Python programs $\mathcal{S}$ that satisfies two primary conditions.

First, the program must be \textit{functionally correct}. This means that for every valid input $x$ from the domain of possible inputs $\mathcal{D}$, the output of the generated program $P(x)$ must match the output of the ideal target function $f_T(x)$ induced by the description $T$. This can be expressed as:
\begin{equation}
    \forall x \in \mathcal{D} : P(x) = f_T(x),
\end{equation}
where $f_T$ is the unknown ground-truth function.

Second, the program must satisfy a set of \textit{formal invariants}, which represent non-negotiable properties required for the solution to be considered valid. These invariants, denoted as a set $I = \{I_1, \dots, I_k\}$, can cover requirements such as \textbf{I/O format}, \textbf{edge-case handling (Completeness)},
\textbf{time/space complexity (Efficiency)}, and \textbf{Python syntax validity (Correctness)}.

%program $P$ that satisfies both the functional correctness criterion (1) and all formal invariants in (2).

\subsection{Formal-Gradient Code Generation (FGCG) Pipeline}

Conventional optimization uses a numerical gradient $\partial \mathcal{L}/\partial \theta$ to steer parameters $\theta$ toward lower loss~$\mathcal{L}$.  In code generation the surface is discrete, so we approximate this signal with a \emph{pseudo--gradient} $g^{(t)}$ obtained by mapping reviewer feedback to an ordered list of required edits follow the previous in Textgrad~\cite{yuksekgonul2024textgrad}:
\begin{equation}
    g^{(t)} \;=\; \operatorname{Parse}\bigl(F^{(t)}\bigr) \;\longrightarrow\; \{(\text{location},\;\text{action})\}_j.
\end{equation}
Each tuple pinpoints a fragment of $P^{(t)}$ together with the type of modification (e.g., ``replace $\mathcal{O}(n^3)$ loop with linear scan'').  Although expressed in natural language, $g^{(t)}$ functions exactly like a real gradient: following its direction increases the likelihood that the next candidate will satisfy the loss template defined below.
As figure~\ref{fig:hook}, at each iteration $t\in\{0,\dots,N\}$ the system maintains a differentiable
text variable $P^{(t)}$ and executes the following four phases:

\begin{enumerate}
    \item \textbf{Forward Generation} ($\mathrm{LM}_{\text{fwd}}$).
          We prompt a code-centric LLM
          $\mathrm{LM}_{\text{fwd}}$  to produce an initial
          or revised program
          \[
              P^{(t)} \;=\; \mathrm{LM}_{\text{fwd}}\!\bigl(\mathcal{T},
              P^{(t-1)}, \nabla P^{(t-1)}\bigr),
          \]
          where $\nabla P^{(t-1)}$—if available—is textual gradient feedback
          from the previous step. Conditioned on the task description~$\mathcal{T}$ and gradient hints from the previous round, the model emits a candidate implementation~$P^{(t)}$.  Intuitively, this component plays the role of a software engineer drafting a solution while being aware of earlier review comments.

    \item \textbf{Backward Evaluation} ($\mathrm{LM}_{\text{bwd}}$).
    A \emph{backward model} reviews the draft.  It returns structured textual feedback
    \mbox{$F^{(t)}{=}[F^{(t)}_{\text{corr}},F^{(t)}_{\text{io}},F^{(t)}_{\text{eff}},F^{(t)}_{\text{comp}}]$},
    where each element analyses the program along a specific axis: Correctness (corr), Input-Output format (io), Efficiency (eff), and Completeness (comp).

    \item \textbf{Gradient Propagation}.
      This step converts every comment, suggestion, or error report into a gradient signal~$\nabla P^{(t)}$ that highlights where and how the code should change.

    \item \textbf{Formal Optimization}.
          We search for an updated program
          $P^{(t+1)}$ that \emph{(i)} descends along $\nabla P^{(t)}$ and
          \emph{(ii)} is accompanied by a proof
          $\Pi^{(t+1)}$ that each invariant holds:
          \begin{equation}
              \Pi^{(t+1)} \;=\;
              \mathrm{LM}_{\text{fwd}}\!\bigl(
              P^{(t)},\nabla P^{(t)},\mathcal{I}\bigr).
          \end{equation}
          The update is accepted if
          $\textsc{Verify}\!\bigl(P^{(t+1)},\Pi^{(t+1)},\mathcal{I}\bigr)=\mathtt{true}$,
          else it is discarded.
          The loop terminates when
$\mathrm{LM}_{\text{bwd}}$ detects no violations
($\forall i: F^{(t)}_{i}\!=\!\text{``pass''}$) or after $N$ iterations.
Formally, each iteration seeks a program
$P^{(t+1)}$ solving the constrained optimization
\begin{equation}
    \min_{P \in \mathcal{S}}
        \; \mathcal{L}\bigl(P;\mathcal{T}\bigr)
    \;\;\text{subject to}\;\;
        \bigwedge_{i=1}^{k} I_i(P)=\mathtt{true},
\end{equation}
where $\mathcal{L}$ is an implicit loss computed by the backward LLM.
CodeGrad approximates the solution by alternating
\emph{gradient proposals} (soft updates) and
\emph{formal verification} (hard constraints),
akin to projected gradient descent in continuous optimization.
\end{enumerate}

\subsection{Experimental Setup}

To evaluate the effectiveness of CodeGrad, we establish a comprehensive experimental setup. We employ \textbf{Qwen 2.5Coder-3B} as our \textit{Forward Model}~\cite{hui2024qwen2}. For the \textit{Backward Model}, we evaluate two distinct models to analyze the impact of critic capability: the lightweight, open-source \textbf{Qwen3-1.7B}~\cite{yang2025qwen3} and the more powerful, proprietary \textbf{GPT4.1-mini}~\cite{OpenAI_GPT4.1}. All open-source models were deployed using the vLLM inference engine on an NVIDIA A6000 GPU~\cite{kwon2023efficient}. For all experiments, we fixed the number of max refinement iterations at two.
We assess performance on two complementary benchmarks:

\begin{itemize}
    \item \textbf{HumanEval} (HE)~\cite{chen2021codex} is a canonical benchmark for functional correctness, comprising 164 human-authored Python problems with rigorous unit tests. We also use its extension, \textbf{HumanEval+}     (HE+)~\cite{liu2023your}, which provides a more comprehensive test suite to better assess solution robustness.
    \item \textbf{LiveCodeBench} (LCB  V6)~\cite{jain2024livecodebench} is a continuously updated benchmark for competitive programming, including problems sourced from active platforms like LeetCode, AtCoder, and Codeforces. We use V6 in its release version, which include 175 questions from Jan 2025 to May 2025.
\end{itemize}

\section{Result}

\iffalse
\begin{table}[h!]
\centering
\caption{
    Performance of the Qwen2.5Coder-3B forward model with various backward models. 
    Values show the absolute score, while the arrow ($\uparrow$) and percentage indicate the performance increase relative to the baseline (NULL backward model).
}
\label{tab:main_single_col}
\begin{tabular}{l ccc}
\toprule
\textbf{Backward Model} & \textbf{HumanEval} & \textbf{HumanEval+} & \textbf{LiveCodeBench V6} \\
\midrule
Null & 0.695 & 0.671 & 0.143 \\
Qwen3-1.7B & \increase{0.787}{13.2} & \increase{0.713}{6.3} & \increase{0.166}{16.1} \\
Qwen3-1.7B (Code) & \increase{0.787}{13.2} & \increase{0.726}{8.2} & \increase{0.160}{11.9} \\
GPT4.1-mini & \increase{0.811}{16.7} & \increase{0.756}{12.7} & \increase{0.251}{75.5} \\
GPT4.1-mini (Code) & \increase{0.884}{27.2} & \increase{0.817}{21.8} & \increase{0.201}{40.6} \\
\bottomrule
\end{tabular}
\end{table}
\fi
\begin{table}[h!]
\centering
\caption{Pass@1 performance across benchmarks. The `+Code` suffix indicates code-enabled backward models. Percentages show increase over the Null backward model baseline.}
\label{tab:backward_model_perf_transposed_grouped}
\footnotesize % Use a smaller font to fit the table
\setlength{\tabcolsep}{4pt} % Adjust column spacing for a compact look
\begin{tabular}{l c cc cc}
\toprule
\multirow{2}{*}{\textbf{Benchmark}} & \multirow{2}{*}{\textbf{Null}} & \multicolumn{2}{c}{\textbf{Qwen3-1.7B}} & \multicolumn{2}{c}{\textbf{GPT4.1-mini}} \\
\cmidrule(lr){3-4} \cmidrule(lr){5-6}
& & Base & +Code & Base & +Code \\
\midrule
HE       & 0.695 & \textbf{0.787}↑13\% & \textbf{0.787}↑13\% & \textbf{0.811}↑17\% & \textbf{0.884}↑27\% \\
HE+      & 0.671 & \textbf{0.713}↑6\%  & \textbf{0.726}↑8\%  & \textbf{0.756}↑13\% & \textbf{0.817}↑22\% \\
LCB V6   & 0.143 & \textbf{0.166}↑16\% & \textbf{0.160}↑12\% & \textbf{0.251}↑76\% & \textbf{0.201}↑41\% \\
\bottomrule
\end{tabular}
\end{table}
\vspace{-0.5cm}
\iffalse
\begin{table*}[h!]
\centering
\caption{Pass@1 by Algorithmic Category, for Forward Model with different Backward model.}
\label{tab:category_perf_revised}
\small % Reduces the font size within the table
\setlength{\tabcolsep}{4pt} % Reduces padding between columns (default is 6pt)
\begin{tabular}{l c cc cc}
\toprule
\multirow{2}{*}{\textbf{Category}} & \multirow{2}{*}{\textbf{Null}} & \multicolumn{2}{c}{\textbf{GPT4.1-mini}} & \multicolumn{2}{c}{\textbf{Qwen3-1.7B}} \\
\cmidrule(lr){3-4} \cmidrule(lr){5-6}
& & Base & +Code & Base & +Code \\
\midrule
Array       & 0.412 & \textbf{0.471}↑14\% & 0.412         & 0.353↓14\% & 0.294↓29\% \\
DP          & 0.000 & \textbf{0.083}      & \textbf{0.083}        & 0.000      & 0.000      \\
Graph       & 0.000 & \textbf{0.062}      & \textbf{0.062}        & \textbf{0.062}      & \textbf{0.125}      \\
Greedy      & 0.000 & \textbf{0.154}      & \textbf{0.154}        & 0.000      & 0.000      \\
Hash Map    & 0.231 & \textbf{0.385}↑67\% & \textbf{0.385}↑67\%  & \textbf{0.308}↑33\% & 0.154↓33\% \\
Math        & 0.119 & \textbf{0.262}↑120\%& \textbf{0.167}↑40\%   & \textbf{0.214}↑80\% & \textbf{0.167}↑40\% \\
Simulation  & 0.071 & \textbf{0.179}↑152\%& \textbf{0.143}↑101\%  & \textbf{0.107}↑51\% & \textbf{0.143}↑101\%\\
String      & 0.333 & \textbf{0.444}↑33\% & \textbf{0.444}↑33\%  & 0.278↓17\% & 0.333      \\
\bottomrule
\end{tabular}
\end{table*}
\fi

\begin{table*}[h!]
\centering
\caption{Pass@1 by Algorithmic Category, for Forward Model with different Backward model.}
\label{tab:category_perf_revised_wide}
\small % Reduces the font size within the table
\setlength{\tabcolsep}{4pt} % Adjusts padding between columns
\begin{tabular}{llcccccccc}
\toprule
\multicolumn{2}{l}{\textbf{Model}} & \textbf{Array} & \textbf{DP} & \textbf{Graph} & \textbf{Greedy} & \textbf{Hash Map} & \textbf{Math} & \textbf{Simulation} & \textbf{String} \\
\midrule
\multicolumn{2}{l}{Null} & 0.412 & 0.000 & 0.000 & 0.000 & 0.231 & 0.119 & 0.071 & 0.333 \\
\midrule
\multirow{2}{*}{\textbf{GPT4.1-mini}} & Base & \textbf{0.471}↑14\% & \textbf{0.083} & \textbf{0.062} & \textbf{0.154} & \textbf{0.385}↑67\% & \textbf{0.262}↑120\% & \textbf{0.179}↑152\% & \textbf{0.444}↑33\% \\
 & +Code & 0.412 & \textbf{0.083} & \textbf{0.062} & \textbf{0.154} & \textbf{0.385}↑67\% & \textbf{0.167}↑40\% & \textbf{0.143}↑101\% & \textbf{0.444}↑33\% \\
\midrule
\multirow{2}{*}{\textbf{Qwen3-1.7B}} & Base & 0.353↓14\% & 0.000 & \textbf{0.062} & 0.000 & \textbf{0.308}↑33\% & \textbf{0.214}↑80\% & \textbf{0.107}↑51\% & 0.278↓17\% \\
 & +Code & 0.294↓29\% & 0.000 & \textbf{0.125} & 0.000 & 0.154↓33\% & \textbf{0.167}↑40\% & \textbf{0.143}↑101\% & 0.333 \\
\bottomrule
\end{tabular}
\vspace{-0.5cm}
\end{table*}

\begin{table}[h!]
\centering
\caption{Pass@1 by Difficulty Level for Forward model with different backward model}
\label{tab:difficulty_perf_revised}
\footnotesize % Use a smaller font size
\setlength{\tabcolsep}{3pt} % Further reduce padding between columns
\begin{tabular}{l c cc cc}
\toprule
\multirow{2}{*}{\textbf{Difficulty}} & \multirow{2}{*}{\textbf{Null}} & \multicolumn{2}{c}{\textbf{Qwen3-1.7B}} & \multicolumn{2}{c}{\textbf{GPT4.1-mini}} \\
\cmidrule(lr){3-4} \cmidrule(lr){5-6}
& & Base & +Code & Base & +Code \\
\midrule
Easy    & 0.558 & 0.558               & 0.558               & \textbf{0.791}↑42\% & \textbf{0.628}↑13\% \\
Medium  & 0.019 & \textbf{0.077}↑305\% & \textbf{0.058}↑205\% & \textbf{0.115}↑505\%& \textbf{0.135}↑611\%\\
Hard    & 0.000 & \textbf{0.013}      & \textbf{0.013}      & \textbf{0.050}      & \textbf{0.025}      \\
\bottomrule
\end{tabular}
\end{table}

\subsection{RQ1: To what extent does the CodeGrad framework improve code generation performance?}

We first establish a baseline using a forward-only configuration with the Qwen2.5Coder-3B model and no backward optimization loop (the "Null" model in Table I). This baseline achieves a pass@1 score of 69.5\% on HumanEval, 67.1\% on HumanEval+, and 14.3\% on the more challenging LiveCodeBench V6. All subsequent results are evaluated relative to this starting point.
Introducing the CodeGrad framework yields substantial improvements. Augmenting the forward model with a lightweight backward model (Qwen3-1.7B) boosts performance by up to 16\% on LiveCodeBench. Employing a more powerful, albeit code-blind, reviewer (GPT4.1-mini) further elevates the scores, with a 17\% absolute increase on HumanEval and a 76\% relative increase on LiveCodeBench.
These collective gains confirm that the CodeGrad framework is an effective mechanism for generating code solutions that are more correct and robust.

\subsection{RQ2: What is the impact of enabling dynamic code execution within the backward model?}

%Qualitative analysis shows that snippet execution surfaces faults invisible to static inspection (e.g., off-by-one errors, zero-division on rare inputs).  The resulting pseudo-gradient pinpoints both the location \emph{and} a counter-example, accelerating convergence. 
Our approach builds upon the established principle of using partial execution for self-debugging, as demonstrated in prior work~\cite{le2022coderl,chen2022codet}, which demonstrates that even minimal, sandboxed probes substantially enhance reviewer effectiveness.
This research question examines the performance difference when the backward model is granted the ability to execute code probes. We compare pairs of architecturally identical reviewers—with and without the sandboxed execution capability (denoted by the \textbf{+Code} suffix in Table II while holding the forward model constant).
The results presented in Table II reveal that the benefit of code execution is highly contingent on the capability of the backward model and the nature of the evaluation benchmark. For the more powerful GPT4.1-mini reviewer, enabling execution on the HumanEval benchmarks yields improvements. The pass@1 score on HumanEval rises from 0.811 to 0.884, a gain of 7.3 percentage points, and on HumanEval+, it increases from 0.756 to 0.817. This demonstrates that a sophisticated reviewer can effectively leverage execution to identify and correct flaws in standard, function-based programming tasks. The effect is less pronounced for the smaller Qwen3-1.7B model, which sees a modest 1.3\% gain on HumanEval+ and no change on the base HumanEval benchmark.

Conversely, the data from Table I shows a performance degradation on LiveCodeBench V6 when execution is enabled. The GPT4.1-mini (Code) configuration scores 5.0\% lower than its non-executing counterpart, and the Qwen3-1.7B (Code) model shows a slight decrease of 0.6\%.
This reversal highlights a key challenge. LiveCodeBench problems involve complex constraints, such as communication via stdin/stdout and strict efficiency requirements, which are difficult to replicate faithfully with isolated code snippets. While execution probes can catch simple runtime errors, they may generate misleading pseudo-gradients in this context. 
%For instance, a probe might confirm the correctness of a logic snippet in isolation, causing the optimizer to focus on it, while failing to detect a larger architectural issue like incorrect time complexity or I/O formatting that causes the full program to fail. Therefore, we conclude that while dynamic execution is a powerful tool for benchmarks like HumanEval, its current implementation can introduce counterproductive signals on more complex, holistic benchmarks like LiveCodeBench, pointing to a critical area for future refinement in probe design and environment emulation.

\subsection{RQ3: How does the performance of CodeGrad vary across different problem categories and difficulty levels?}
To understand the conditions under which CodeGrad is most beneficial, we analyze its performance on problems grouped by algorithmic category (Table II) and difficulty (Table III). We analyze the problems from the LiveCodeBench dataset along two dimensions: their predefined difficulty level and their algorithmic category, which we assign based on the classification from LeetCode~\cite{leetcode_problemset}.
The breakdown by category reveals that the framework's advantages are most pronounced in complex algorithmic domains. For categories like Dynamic Programming (DP), Graph, and Greedy, the baseline model fails to produce a single correct solution (0.0\% pass@1). In contrast, the full CodeGrad approach successfully solves problems in these areas, demonstrating its ability to navigate complex logic and constraints. The largest relative gains are observed in Simulation (+101\%), Math (+40\%), and Hash map (+66\%), where iterative refinement and formal checks are critical for correctness.

The analysis by difficulty level in Table III reinforces this finding. While CodeGrad offers a respectable 12.5\% improvement on "Easy" problems, its impact is transformative for more challenging tasks. It boosts performance on "Medium" difficulty problems by over 610\% relative to the baseline and successfully generates correct solutions for "Hard" problems where the baseline fails completely. 
%The CodeGrad is indispensable for tackling problems that require deep reasoning, multi-step logic, and robust handling of edge cases.

\section{Implication and Future Work}

%The primary contribution of this research is using the formal specification check to guide the optimization of LLM iteration process. The approach has potential to first improve reliability of AI-generated code second, in a broader way, using the formal check in to enhance the interprebility of LLM
%This approach has the potential to significantly the reliability of AI-generated code, making it more suitable for deployment in high-stakes domains such as competitive programming, scientific computing, and safety-critical systems where correctness is non-negotiable.

While our results demonstrate the effectiveness of the CodeGrad framework, there are several directions for future exploration. 
\textbf{First}, we will enhance the efficiency of the iterative refinement process. We plan to replace the current fixed-iteration method with a dynamic, search-based approach that identifies the most cost-effective path to an optimal solution. 
%we plan to enhance the efficiency of the iterative refinement process itself. The current implementation relies on a fixed number of iterations, which can be suboptimal. Future work will investigate adaptive termination criteria, such as halting when a solution's proof stabilizes or when refinement yields diminishing returns.
\textbf{Second}, we aim to translate our research prototype into a practical, developer-centric tool. This involves designing a robust framework that enables software engineers to directly apply CodeGrad to their own development workflows, potentially as an integrated development environment (IDE) plugin or a standalone library. \textbf{Finally}, we will explore extending the core principles of CodeGrad beyond code generation. The concept of using formal feedback as a textual pseudo-gradient is highly generalizable. We plan to investigate its application in related domains such as automated program repair, where formal invariants could guide the fixing of bugs, and interactive proof generation, where the framework could help refine mathematical or logical arguments expressed in natural language.

%First, the efficiency of the iterative refinement process could be enhanced. The current loop relies on a fixed number of iterations, which may be suboptimal. Future work could investigate adaptive termination criteria or employ more search algorithms within the to find valid solutions more quickly. Furthermore, we will design a framework with our complete approach in the future, which can enable developers to directly use our approach in downstream task in software engineering. Also we will try more task like .... in the future not only limited to code generation.

\section{Conclusion}
This work introduced CodeGrad, a framework that successfully integrates gradient-based iterative refinement with formal verification for LLM-based code generation. Our results demonstrate that this hybrid approach enhances performance, achieving substantial gains on challenging benchmarks like LiveCodeBench. The framework proves particularly effective for complex problems requiring deep reasoning and strict constraint adherence, demonstrating a viable path toward generating more reliable, robust, and formally justified AI-assisted software for high-stakes applications.

\bibliographystyle{IEEEtran}
\bibliography{software,actr}% common bib file
\end{document}